%% file: main.tex
\documentclass[journal]{IEEEtran}

\IEEEoverridecommandlockouts                            

\usepackage[pdftex]{graphicx}
\graphicspath{{figure/}}
\usepackage{amsmath}
\usepackage{amssymb}
\usepackage{url}

\usepackage{amsthm} 
\usepackage{cite}
\usepackage{color}
\usepackage{comment}
\usepackage{fancyhdr} 

\usepackage{algorithm}
\usepackage{algorithmic}
\usepackage{bm}
\usepackage{bbm}

\newtheorem{remark}{\textbf{Remark}}

\newcommand{\bx}{\bm{x}}
\newcommand{\bth}{\bm{\theta}}

\newcommand{\by}{\bm{y}}
\newcommand{\bz}{\bm{z}}

\newcommand{\bw}{\bm{w}}

\newcommand{\bu}{\bm{u}}
\newcommand{\bl}{\bm{\lambda}}

\newcommand{\sN}{\mathcal{N}}

\newcommand{\bmu}{\bm{\mu}}

\newcommand{\nonl}{\renewcommand{\nl}{\let\nl\oldnl}}
\newboolean{showcomments}
\setboolean{showcomments}{true}
\newcommand{\lina}[1]{  \ifthenelse{\boolean{showcomments}}
	{ \textcolor{red}{(Lina says:  #1)}} {}  }
\newcommand{\zhaolin}[1]{  \ifthenelse{\boolean{showcomments}}
	{ \textcolor{blue}{(Zhaolin says:  #1)}} {}  }

\allowdisplaybreaks 

\title{  Online Learning and  Distributed Control for Residential Demand Response
}

\author{Xin Chen,~\IEEEmembership{Student Member,~IEEE,} Yingying Li, Jun Shimada,  Na Li,~\IEEEmembership{Member,~IEEE}
	\thanks{  X. Chen, Y. Li and N. Li are with the School of Engineering and Applied Sciences, Harvard University, USA. Email: (chen\_xin@g.harvard.edu, yingyingli@g.harvard.edu,  nali@seas.harvard.edu). 
	}
	\thanks{J.  Shimada is with ThinkEco Inc., New York, USA. Email: jun@thinkecoinc.com.}
	\thanks{ 
The work was supported by 
NSF CAREER: ECCS-1553407 and
NSF EAGER: ECCS-1839632.} 
}

\begin{document}

\maketitle

\pagestyle{plain}
	
\input{Abstract}

\input{Introduction}

\input{Preliminaries}

\input{Problem}

\input{Algorithm}

\input{Simulation}

\input{Conclusion}

\input{Appendix1}

\input{Reference}
\end{document}

%% file: Abstract.tex
\begin{abstract}
This paper studies the automated control method for regulating air conditioner (AC) loads in incentive-based residential demand response (DR). The critical challenge is that the customer responses to load adjustment are uncertain and unknown in practice. In this paper, we formulate the AC control problem in a DR event as a  multi-period stochastic optimization that integrates the indoor thermal dynamics and customer opt-out status transition. Specifically, machine learning techniques including
Gaussian process and logistic regression are employed to learn the unknown thermal dynamics model and customer opt-out behavior model, respectively. We consider two typical DR objectives for AC load control: 1) minimizing the total demand, 2) closely tracking a regulated power trajectory. Based on the Thompson sampling framework,  we propose an online DR control algorithm to learn customer behaviors and make real-time AC control schemes. This  algorithm considers  the influence of various environmental factors on  customer behaviors and is implemented in a distributed fashion to preserve the  privacy of customers.
Numerical simulations  demonstrate the control optimality and learning efficiency of the proposed algorithm.

\end{abstract}


%% file: Introduction.tex
\section{Introduction} \label{sec:introduction}

\IEEEPARstart{D}{ue} to increasing 
 renewable generation and growing peak load,  electric power systems 
are inclined to confront a deficiency of reserve capacity.  As a typical example, in
 mid-August 2019, Texas grid 
 experienced  record electricity demand and severe  reserve emergency that were caused by the heat wave and reduced wind generation.  The electricity price once soared to 9\$/kWh and the
Electric Reliability Council of Texas (ERCOT) issued Level 1 Energy Emergency Alert to call  upon voluntary energy conservation and all available generation sources \cite{DRass_2019}. To cope with such problems,  demand response (DR)    is an economical and  sustainable solution that strategically
motivates load adjustment from end users to meet the needs of power supply \cite{usdep}.
In particular, residential loads  account for a large
share of the total electricity usage (e.g. about $38\%$ in the
U.S. \cite{eleuseUS}), 
which can release significant power flexibility to facilitate  system operation
through a coordinated dispatch. 
Moreover, the widespread deployment of advanced meters, smart plugs, and  built-in controllers
enables
the remote monitoring and control  of  electric appliances
with  two-way communications between households and  load service entities
(LSEs).
This  makes it technically feasible to implement residential DR, and well-designed DR control algorithms are necessitated  to fully exploit potential flexibility.

The mechanisms for residential DR are mainly categorized as ``price-based"   and ``incentive-based" \cite{drreview}. The price-based DR programs \cite{price1} use various pricing schemes, such as time-of-use pricing, critical peak pricing, and real-time pricing, to influence and guide the  electricity usage. 
In  incentive-based DR programs  \cite{ince2,ince3}, the LSEs
recruit customers to adjust their load demands   in  DR events  with financial incentives, e.g. cash, coupon, raffle, rebate, etc.  A typical residential DR event consists of two periods:
1) the \emph{preparation period} when the LSEs anticipate an upcoming load peak or system emergency   and call upon  customers to participate  with incentives (day-ahead or hours-ahead); 
2) the \emph{load adjustment period} when the electric appliances of  participating customers are  controlled to achieve certain DR goals. 
Meanwhile, customers are allowed to opt out (e.g.  by clicking the “opt out” button in the smart phone app) if unsatisfied and override the control commands from the LSEs. 
In the end, the LSEs 
pay   customers according to their actual contributions. 
 In practice, the load adjustment period usually lasts for a  few hours, and the control cycle of electric appliances varies from 5 to 30 minutes depending on the actual DR setting, which is enabled by the advanced meters with second or minute  sampling rates.
This paper focuses on the real-time  control  of electric appliances \emph{during the load adjustment period} 
from the perspective of LSEs.


According to the investigations in \cite{survey1, survey2},  offering the override (opt-out) option can greatly  enhance the customers' acceptance of direct load control, which is even more  effective than financial incentives.  Hence, customers are generally 
 authorized to have the opt-out option 
in modern residential DR programs.
However, the customer opt-out behaviors are uncertain and unknown to  LSEs in practice, which brings significant challenges to  the  real-time  DR control. References \cite{survey1, survey2, survey3,survey4} indicate that  customer DR behaviors 
are influenced by \emph{individual preference} and \emph{environmental factors}.
The individual preference  relates to customers' intrinsic socio-demographic characteristics, e.g.   income, education, age,   attitude to energy saving, etc. The environmental factors refer to  real-time  externalities  such  as
electricity price,  
indoor  temperature,  offered incentive, weather conditions,  etc. 
Without considering customers'  actual  willingness, a blind DR control scheme may lead to high opt-out rates and inefficient load adjustment.

To address the uncertainty issue, 
data-driven learning techniques can be 
employed to learn  customer DR behaviors with historical data and 
 through online interactions and observations. 
Comprehensive reviews on the application of reinforcement learning (RL) for DR are provided in  \cite{rldrre,rlreview}.
References \cite{ems1,ems2, ems3} 
design home energy management systems to optimally schedule electric appliances under time-varying electricity price, where $Q$-learning is used to learn  customer preferences and make rescheduling decisions. In \cite{realdr}, a 
 real-time DR strategy is presented for optimal arrangement of home appliances based on deep RL and  policy search algorithms, considering the
uncertainty of the resident’s behavior,  electricity price,
and outdoor temperature.
  Reference \cite{tclbat} applies batch RL  to dispatch  thermostatically controlled loads (TCLs) with the exploitation of exogenous data for  day-ahead scheduling. 
 Reference \cite{incrl} proposes 
  an incentive-based DR algorithm
 using RL and deep neural networks, to assist LSEs in designing the optimal incentive rates with
uncertain energy  prices and demands.
  In
   \cite{phase0-1,phase0-2}, the multi-armed bandit  method and its variants are adopted to  select the  right customers for DR participation at the preparation  period to deal with
    unknown customer responses.
 However, 
 for real-time  load control in incentive-based DR,
 most existing works   do not  consider or overly simplify the uncertainty of customer behaviors, and  the influence of various environmental factors is generally neglected. 
 This causes significant mismatches between theory and practice. Hence,
 the development of real-time DR control algorithms that take into account customer opt-out behaviors  remains largely unresolved.
 
 
\textit{Contribution.} This paper studies the incentive-based residential DR programs that control air conditioner (AC) loads to optimize certain DR performances, e.g., minimizing the total AC load or closely tracking a target power trajectory. 
 To this end, we propose a novel framework to model the real-time DR control as a multi-period stochastic optimization problem that integrates thermal dynamics and customer behavior transitions. 
 In particular,  Gaussian process (GP) \cite{gp_ml} is adopted to build a non-parametric indoor thermal dynamical model from historical metering data, and  
 logistic  regression  \cite{logis} is used to model the customer opt-out behaviors under  the influence of   environmental factors. 
 Based on the Thompson sampling (TS) framework \cite{ts1},  we develop a distributed online DR control algorithm to 
 learn the customer opt-out behaviors and make real-time AC power control decisions.
 The main merits of  the proposed  algorithm are summarized as follows:
 \begin{itemize}
       \item [1)] The individual preferences of customers and  time-varying environmental factors are taken into account, which  improves the predictions on customer opt-out behaviors and leads to efficient AC control schemes.
     \item [2)] This algorithm is implemented in a distributed manner and thus can be directly embedded in local household AC appliances, smart plugs, or smart phone apps. Moreover, the communication burdens are mitigated and  the customer privacy can be preserved.
     \item [3)] Inheriting the merits of TS, this algorithm has a convenient decomposition structure of learning and optimization, and 
     strikes an effective  balance between exploration and  exploitation in the online learning  process. 
 \end{itemize}

The remainder of this paper is organized as follows: Section \ref{sec:preliminary} provides a preliminary introduction on GP and TS. Section \ref{sec:problem} presents the optimal AC control models with the learning techniques. Section \ref{sec:algorithm} develops the distributed online learning and  AC control algorithm. 
 Numerical tests are carried out  in Section \ref{sec:simulation}, and conclusions are drawn in Section \ref{sec:conclusion}.

%% file: Preliminaries.tex
\section{Preliminaries on Learning Techniques}  \label{sec:preliminary}

This section provides a preliminary introduction on  the two key learning techniques used in this paper, i.e., Gaussian process  and Thompson sampling.

\subsection{Gaussian Process}

Gaussian process  is a  non-parametric supervised  machine learning method \cite{gp_ml} that has  been widely used to model nonlinear system dynamics \cite{gp_sys}. A formal definition of GP over a function  $f(\bx)$ is that any finite number of function realizations 
$(f(\bx_1), f(\bx_2),f(\bx_3),\cdots)$ are random variables and follow a joint Gaussian distribution, which is fully  specified by the mean function $m(\bx)$ and the (kernel) covariance function $k(\bx,\bx')$. 
Consider learning an unknown function $y \!=\! f(\bx)$ based on a training dataset $\mathcal{D}$ of $n$ noisy observations, i.e. $\mathcal{D}:=\{(\bx_i, \hat{y}_i) |\, i\!=\!1,\cdots,n\}$. It aims to infer the function value $f(\bx_*)$ for a new  point $\bx_*$.
Denote  $\hat{\by}_\mathcal{D}\!:=\! (\hat{y}_i)_{i=1}^n$ and
$\bm{f}_\mathcal{D}\!:=\! (f(\bx_i))_{i=1}^n$.
By the GP definition, $(\bm{f}_\mathcal{D}, f(\bx_*))$ are assumed to be random variables and follow a joint Gaussian distribution
\begin{align*}
    \begin{bmatrix}
   \bm{f}_\mathcal{D}\\ f(\bx_*)
    \end{bmatrix} \sim \mathcal{N}  \begin{pmatrix}
    \begin{bmatrix}
    \bm{m}_\mathcal{D}\\ m(\bx_*)
    \end{bmatrix}
    ,\, \begin{bmatrix}
    K_{\mathcal{D},\mathcal{D}} & \bm{k}_{*,\mathcal{D}}\\
    \bm{k}_{*,\mathcal{D}}^\top & k(\bx_*,\bx_*)
    \end{bmatrix}
     \end{pmatrix},
\end{align*}
where vector
 $\bm{k}_{*, \mathcal{D}}\!:=\! [k(\bx_*,\bx_1), \cdots, k(\bx_*,\bx_n)]^\top$ and $\bm{m}_\mathcal{D}\!:=\! (m(\bx_i))_{i=1}^n$. $K_{\mathcal{D},\mathcal{D}}\!\in\!
\mathbb{R}^{n\times n}$ is the covariance matrix, whose $ij$-component is $ k(\bx_i,\bx_j)$. Conditioning on the given observations $\mathcal{D}$, it is known that the posterior distribution of $f(\bx_*)|(\bm{f}_\mathcal{D}\!=\!\hat{\by}_\mathcal{D})$ is also  Gaussian, i.e. $\mathcal{N}(\mu_{*|\mathcal{D}}, \sigma^2_{*|\mathcal{D}} )$, with  the closed form
\begin{subequations} \label{eq:GP}
\begin{align}
    \mu_{*|\mathcal{D}}& = m(\bx_*) +\bm{k}_{*, \mathcal{D}}^\top\, K_{\mathcal{D},\mathcal{D}}^{-1}\,(\hat{\by}_\mathcal{D}-\bm{m}_\mathcal{D}),\label{eq:GP:mean}\\
    \sigma^2_{*|\mathcal{D}} & = k(\bx_*,\bx_*) - \bm{k}_{*, \mathcal{D}}^\top\, K_{\mathcal{D},\mathcal{D}}^{-1}\,\bm{k}_{*, \mathcal{D}}.\label{eq:GP:var}
\end{align}
\end{subequations}
Then the mean value $\mu_{*|\mathcal{D}}$ (\ref{eq:GP:mean}) can be used as the prediction on $f(\bx_*)$, and the variance $ \sigma^2_{*|\mathcal{D}}$ (\ref{eq:GP:var})  provides a confidence estimate for this prediction. The merits of GP include 1) GP is a non-parametric method that avoids the bias of model selection; 2) GP works well with small datasets; 3) GP can incorporate prior domain knowledge by defining priors on hyperparameters or using a particular covariance function.
 The major issue of GP is  the computational complexity, which  scales cubically in the number of observations, i.e., ${O}(n^3)$.

\subsection{Thompson Sampling}

Thompson sampling   \cite{ts1} is a prominent Bayesian learning framework that was originally developed to solve the multi-armed bandit (MAB)  problem \cite{mab} and   can  be extended to tackle other online learning problems.
Consider a classical $T$-periods MAB problem where an agent selects an  action (called ``arm")  $a_t$ from the action set $\mathcal{A}$ at each time $t \in \{1,2,\cdots,T\}$. After taking  action $a_t$, the agent observes an outcome $z_t$ that is randomly generated
from a conditional  distribution $\mathcal{P}_\theta(\cdot|a_t)$, and then  obtains a reward $r_t = r(z_t)$ with known reward  function $r(\cdot)$. 
The agent is initially uncertain about the parameter $\theta$ in  $\mathcal{P}_\theta$ but 
aims to maximize the total expected reward using the observation feedback.
To achieve good performance, it 
 is generally required to take actions with an effective balance between 1)
 \emph{exploring} poorly-understood actions  to  gather
new information that may improve future reward  and 2) \emph{exploiting}  what is known for decision to maximize the immediate reward.

TS is a straightforward online learning algorithm that strikes an effective balance between exploration and exploitation.
 As shown in Algorithm \ref{tsalg},
TS treats the unknown $\theta$ as a random variable and
represents the initial belief on  $\theta$ with a prior distribution $\mathcal{P}$. At each time $t$, TS draws  a random sample $\hat{\theta}$ from $\mathcal{P}$ (Step \ref{ln:draw}), then takes the optimal action that maximizes the expected reward based on the sample $\hat{\theta}$ (Step \ref{ln:opt}). After  outcome $z_t$ is observed, the Bayesian rule is applied to update the distribution $\mathcal{P}$ over $\theta$ and obtain the posterior (Step \ref{ln:updat}). 

\begin{algorithm}
 \caption{Thompson Sampling (TS) Algorithm \cite{ts1}}
 \begin{algorithmic}[1]  \label{tsalg}
   \STATE \textbf{Input:} Prior distribution $\mathcal{P}$ on $\theta$. 
  \FOR {$t = 1$ to $T$}
  
  \STATE Sample $\hat{\theta}\sim \mathcal{P}$. \label{ln:draw}
  \STATE $a_t \leftarrow \arg\max_{a\in \mathcal{A}} \mathbb{E}_{\mathcal{P}_{\hat{\theta}}}[r(z_t)|a_t = a] $.  \label{ln:opt}
  
  Apply $a_t$ and observe $z_t$.
  
  \STATE Posterior  update: 
$ \ \   \mathcal{P}\leftarrow \frac{\mathcal{P}(\theta) \mathcal{P}_\theta (z_t|a_t)}{\int_{\tilde{\theta}} \mathcal{P}(\tilde{\theta}) \mathcal{P}_{\tilde{\theta} }(z_t|a_t)\, d\tilde{\theta}}.
$ \label{ln:updat}
  \ENDFOR 
 \end{algorithmic} 
 \end{algorithm}
 
The main features of the TS algorithm are listed below:
\begin{itemize}
   \item  As outcomes accumulate, the predefined prior distribution will be washed out and the posterior converges to the true distribution or concentrates on the true value of $\theta$.
    \item  The TS algorithm encourages exploration by the random sampling (Step \ref{ln:draw}). As the posterior distribution gradually concentrates, less exploration and more exploitation will be performed, which leads to an effective balance. 
    \item  The key advantage of the TS algorithm is that  a complex online problem is decomposed into a Bayesian learning task (Step \ref{ln:updat}) and an offline optimization task (Step \ref{ln:opt}) \cite{ts2}, while the  optimization remains the original  formulation without being complicated by the learning task.
\end{itemize}


%% file: Problem.tex
\section{Problem Formulation} \label{sec:problem}

Consider the residential DR program that controls  AC power consumption for load adjustment, where  a system aggregator (SA) interacts with $N$ residential customers over  sequential DR events.  Each DR event\footnote{To avoid confusion,  
in the following text, we  specifically refer to the load adjustment period  when referring to ``DR event".}  is formulated as a finite time horizon $[T]:=\{1,2,\cdots,T\}$ with the time gap $\Delta t$.
Depending on the practical AC control cycle, $\Delta t$ may be different  (e.g., 5 minutes or 10 minutes) across DR events\footnote{Generally, the control time gap $\Delta t$ should be larger than the  metering data collection period, which is enabled by  current smart meters with  second or minute sampling rates.}, and the time length $T$ could also be different (e.g., 2 hours or 3 hours). 
The SA aims to learn customers' opt-out behaviors and make  real-time AC control decisions to optimize aggregate DR performance. 
 In this paper, we focus on the control of household mini-split AC units, while 
the proposed method is  applicable to heating, ventilation and
air conditioning (HVAC) systems, space heaters, and other TCLs. Moreover, the proposed method works for both the AC heating and cooling cases separately.

In this section, we firstly establish the AC control model for each individual customer during a DR event, then use Gaussian process and logistic regression to  learn the unknown  thermal dynamics and customer opt-out behaviors, respectively. With all these in place, two system-level optimization models with different DR objectives are built for the SA to generate optimal AC power control schemes over $N$ customers.

\subsection{Customer-Side AC Control Model During A DR Event}

\subsubsection{Decision Variable}
During a DR event, 
denote $u_{i,t}$ as the AC power consumption for
customer $i\!\in\! [N]\!:=\!\{1,\cdots,N\}$ at time $t\!\in\! [T]$, which is the \emph{decision variable} and  satisfies
\begin{align} \label{eq:ucon}
    0\leq u_{i,t}\leq  \bar{u}_i,\ |u_{i,t}\!-\!u_{i,t-1}|\leq \Delta u_i, \quad \forall t\in[T] 
\end{align}
where $\bar{u}_i$ is the   rated AC power capacity and $\Delta u_i$ denotes the  AC power drift limit that prevents dramatic  changes.  In \eqref{eq:ucon},  $u_{i,t}$ is  continuously controllable, since it indeed denotes  the average AC power during the time interval $[ t\!-\!1,t]$, which can be  realized by appropriately adjusting the AC cycling rate, i.e., the time ratio of the on status over $\Delta t$. Nevertheless, a discrete AC control model for customer $i$ with 
\begin{align*}
    \qquad  u_{i,t} = \delta_{i,t}\cdot\bar{u}_i,\ \delta_{i,t}\in\{0,1\}, \quad \forall t\in[T]
\end{align*} 
 can  be applied as well, where the AC unit  switches between the on ($\delta_{i,t}=1$) and off ($\delta_{i,t}=0$) modes. 


\subsubsection{Opt-Out Status and Transition}
Denote binary variable $z_{i,t}\!\in\!\{0,1\}$ as the 
\emph{opt-out status} of customer $i$ at time $t$, which equals $1$ if  customer $i$ stays in and $0$ if  opts out. Initialize
 $z_{i,0}\! = \!1$  for all  customers 
at the beginning of a DR event. 
As mentioned above,  customer opt-out behaviors are influenced by  various environmental factors. 
Accordingly, the binary opt-out statuses $\bm{z}_i\!:=\!(z_{i,t})_{t\in[T]}$ of  each customer $i\in[N]$ are modelled as a time series of random variables
that are independent of other customers and follow the  transition probability (\ref{eq:condi}): 
         \begin{subequations} \label{eq:condi}
            \begin{align}
    \mathbb{P}(z_{i,t} \!=\! 0|z_{i,t-1} \!=\!0)& =1,& \forall t\in [T] \label{eq:condi:0}\\
\mathbb{P}( z_{i,t}\!=\! 1|{z_{i,t-1}} \!=\!1)& = p_i( \bw_{i,t}), & \forall t\in [T]\label{eq:condi:1}
    \end{align}
    \end{subequations}
Here, $\bw_{i,t}$ is the column vector that collects the environmental factors at time $t$, which will be elaborated in the next part.
 $p_i( \bw_{i,t})\in[0,1]$ is the transition probability function that captures how environmental factors  $\bw_{i,t}$ affect the customer opt-out behaviors. 
\begin{figure}[thpb]
	\centering
		\includegraphics[scale=0.3]{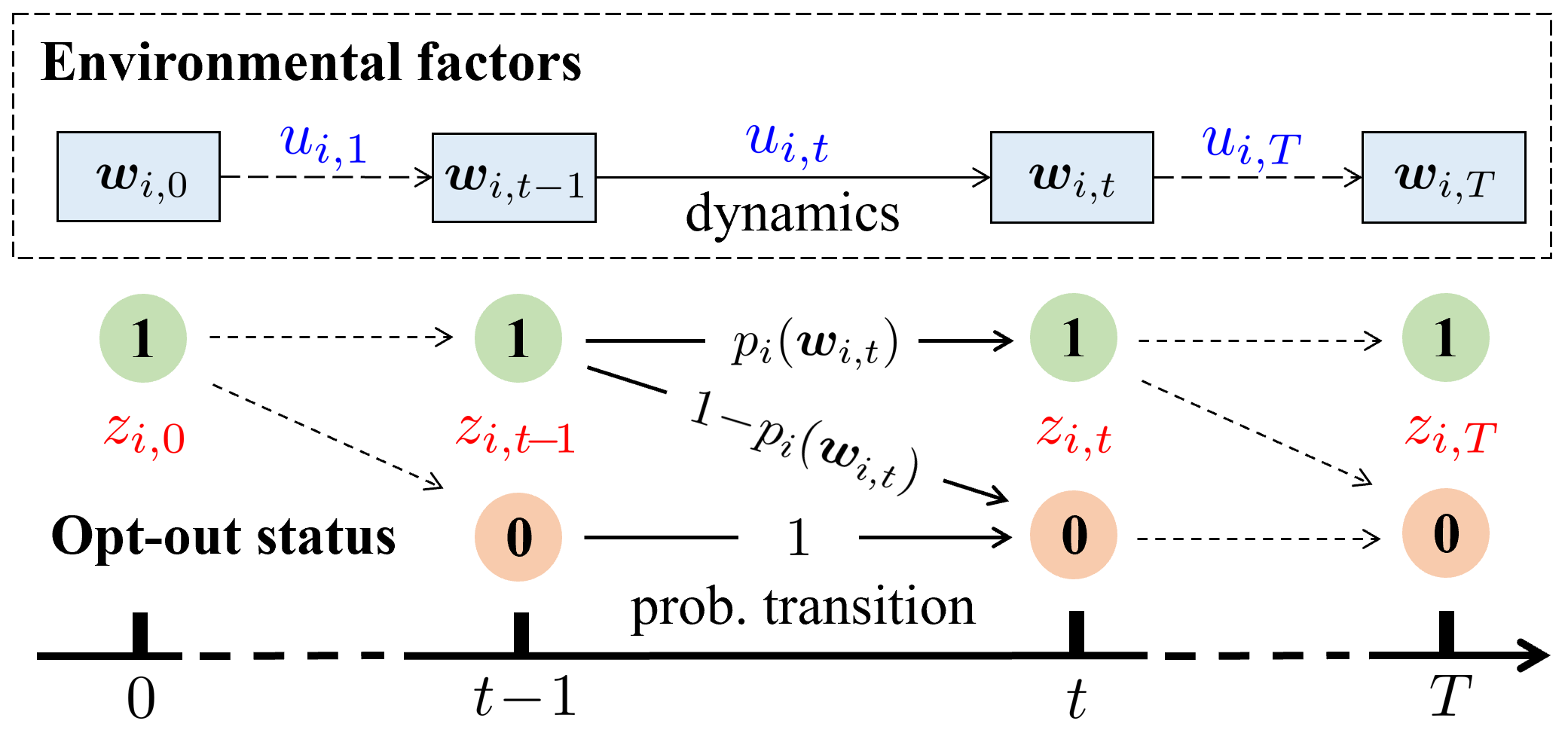}
	\caption{Opt-out status transition and dynamics of AC control in a DR event.
	}
	\label{fig:dyna}
\end{figure}

As illustrated in Figure \ref{fig:dyna}, equation (\ref{eq:condi:0}) enforces the DR rule that
once customer $i$ opts out at a certain time, this customer will remain the opt-out status  for the rest of the current DR event. 
Equation (\ref{eq:condi:1}) indicates that if customer $i$ stays in at time $t\!-\!1$, this customer may remain  stay-in at next time $t$ with  probability $p_i( \bw_{i,t})$, or choose to opt out 
 with probability $1-p_i( \bw_{i,t})$. We further explain the transition model \eqref{eq:condi} with the following remark.

\begin{remark}
\normalfont 
The opt-out status transition model \eqref{eq:condi} exhibits the Markov property, where the 
transition probability $p_i(\bw_{i,t})$ is  functional on the environmental factors $\bw_{i,t}$ at time $t$. This facilitates the subsequent development and solution of the optimal AC control models, but 
does not sacrifice the modeling  generality. Because it is free to choose suitable environmental factors, so that  all the useful information is captured in $\bw_{i,t}$ and the Markov property is preserved.
Basically, by including all the necessary known information in the enlarged state at time $t$, which is known as  \emph{state augmentation}, any multi-period control problem can generally be modelled  as a Markov decision process \cite{markov}. See the next part for the detailed definition and selection of the environmental factors.

\end{remark}

\subsubsection{Environmental Factors} 

  Based on the empirical investigations in \cite{survey1,survey2,survey3,survey4}, we present below several key environmental factors that influence customers' opt-out behaviors. In particular, the first three factors are affected by the AC control scheme, thus their dynamics models are introduced as well. 


$\bullet$ (\emph{Indoor Temperature}). Denote $s_{i,t}$ as the indoor temperature  for customer $i$ at time $t$. The associated thermal dynamical model can be formulated as 
\begin{align}\label{eq:therm}
  s_{i,t} = f_i\left( s_{i,t-1}, s_{i,t-1}^{\mathrm{out}}, u_{i,t}  \right),\quad   t\in[T]
\end{align}
where $s_{i,t}^{\mathrm{out}}$ is the outdoor temperature at time $t$, $(s_{i,0}, s_{i,0}^{\mathrm{out}})$ are the initial temperatures in the beginning of the DR event, and $f_i(\cdot)$ denotes the thermal dynamics function. 

$\bullet$ (\emph{Accumulated Thermal Discomfort}). We define $d_{i,t}$ as the accumulated thermal discomfort for customer $i$ at time $t$, and let it follow the dynamics \eqref{eq:discom} with $d_{i,0} =0$:
 \begin{align} \label{eq:discom}
    d_{i,t} = d_{i,t-1} +  \Delta t\cdot\! \big(\!\max\!\big\{ s_{i,t}-s^{\mathrm{set}}_i,\,0 \big\}\big)^2,\quad  t\in[T]
\end{align}
where $s_i^\mathrm{set}$ denotes the defaulted  AC setting temperature by  customer $i$. The  operator $\max\{x,0\}$ takes the larger value between $x$ and $0$, which 
means  that only the indoor temperature higher than the setpoint will
cause thermal discomfort for the  AC cooling  case in the summer. Besides, the quadratic form in (\ref{eq:discom})  captures  that the thermal discomfort increases  faster as the temperature deviation becomes larger \cite{Liu_behavior_2019}.  

$\bullet$ (\emph{Incentive Credit}). Denote $r_{i,t}$ as the  incentive credit offered to customer $i$ at time $t$. We consider a general incentive scheme \eqref{eq:reward},  while other incentive  schemes can  be used as well.
\begin{align} \label{eq:reward}
    r_{i,t} = r_0 + r_1\cdot \bar{u}_i\cdot t 
    +r_2 \cdot\Delta t\sum_{\tau=1}^t |u_{i,\tau}^{\mathrm{set}}\! -\!  u_{i,\tau}|,\ t\in[T].\!
\end{align}
In \eqref{eq:reward}, the first term $r_0$ is the base credit for DR participation. The second term is the stay-in bonus that is proportional to time $t$ 
and the AC power capacity $\bar{u}_i$ with   coefficient $r_1$.
The third term is the reimbursement for the actual load adjustment with credit coefficient $r_2$,  where $u_{i,t}^\mathrm{set}$ is the associated AC power at time $t$ to maintain the setting temperature  $s_i^{\mathrm{set}}$.

Essentially, $(u_{i,t}^{\mathrm{set}})_{t\in[T]}$ can be regarded as the baseline AC power consumption of customer $i$  when no DR control is implemented. With the thermal dynamical model \eqref{eq:therm} and given $(s_{i}^{\mathrm{set}},s_{i,t-1}^{\mathrm{out}} )$, 
one can compute $u_{i,t}^{\mathrm{set}}$ by solving equation \eqref{eq:solveuset}:
\begin{align}\label{eq:solveuset}
    s_{i}^{\mathrm{set}} - f_i( s_{i}^{\mathrm{set}}, s_{i,t-1}^{\mathrm{out}}, u_{i,t}^{\mathrm{set}} ) =0,
\end{align}
which follows the definition of maintaining the defaulted setting indoor  temperature $s_i^{\mathrm{set}}$.




Other key environmental factors that would influence customers' opt-out  behaviors include
real-time electricity prices,  weather conditions,  duration of the DR event, fatigue effect, etc. 
 These factors are  treated as given parameters that can be obtained or  predicted  ahead  of time. Accordingly, the vector $\bw_{i,t}$ in \eqref{eq:condi:1} can be defined as the combination of the environmental factors mentioned above:
 \begin{align}\label{eq:wdef}
     \bw_{i,t}:=(s_{i,t}, d_{i,t}, r_{i,t}, \text{electricity price at }t,\,\cdots).
 \end{align}

\begin{remark} 
\normalfont We note that the definition and selection of useful environment factors are complex and tricky in practice. For instance, the definition of $\bw_{i,t}$ in \eqref{eq:wdef} only contains the present status at time $t$, while the past values and  future predictions may also be included in $\bw_{i,t}$  to capture the temporal dependence. This  is related to the feature engineering problem in machine learning, which is expected to be  conducted based on real data and make
 a trade-off between complexity and effectiveness. Nevertheless, the proposed learning and AC control method is a general framework that is applicable to different choices of the environmental factors.
\end{remark}

One critical issue for the residential AC control is that  the thermal dynamics function $f_i(\cdot)$ in (\ref{eq:therm}) and the customer opt-out behavior function $p_i(\cdot)$ in (\ref{eq:condi:1}) are generally unknown. 
 To address this issue, learning techniques are used to estimate the unknown models with real data,  which are presented in the following two subsections.  

\subsection{Learning for  Thermal Dynamics Model}

The practical implementation of AC control for residential DR is generally achieved through  smart plugs or smart AC units with built-in controllers and sensors. These smart devices are able to measure, store, and communicate the  temperature and AC power data in real time. Hence,  
 the thermal dynamics model (\ref{eq:therm}) can be estimated based on fine-grained historical measurement data.
To this end, we provide the following two  thermal model estimation schemes.

\subsubsection{Linear  Model} Given a time series of  historical indoor/outdoor temperature and AC power data, one can fit a classical linear  thermal dynamics model (\ref{eq:lineartherm}) \cite{lina_dr}  and obtain the coefficients $(\kappa_i,\eta_i)$ via linear regression:
 \begin{align}\label{eq:lineartherm}
     s_{i,t} = s_{i,t-1} + \kappa_i\cdot(s_{i,t-1}^{\mathrm{out}}-s_{i,t-1} ) + \eta_i\cdot u_{i,t},
 \end{align}
 where coefficients $\kappa_i$ and $\eta_i$  specify the thermal characteristics of the room with AC and the ambient. A positive (negative)  $\eta_i$ indicates that AC works in the heating (cooling) mode.

\subsubsection{Gaussian Process Model} 
An alternative scheme is to employ the Gaussian process method (introduced in Section \ref{sec:preliminary})  to model 
 the thermal dynamics as \eqref{eq:gpthermal}, 
  which can capture the nonlinearity in the data pattern: 
  \begin{align} \label{eq:gpthermal}
     s_{i,t} =  m(\bx_{i,t-1}) +\bm{k}_{*, \mathcal{D}}(\bx_{i,t-1})^\top K_{\mathcal{D},\mathcal{D}}^{-1}\,(\hat{\by}_\mathcal{D}-\bm{m}_\mathcal{D}),
 \end{align}
where $\bx_{i,t-1}\!:=\!(s_{i,t-1}, s_{i,t-1}^{\mathrm{out}}, u_{i,t} )$, and the notations with subscript $\mathcal{D}$ denote the corresponding terms associated with the historical (training) dataset as presented in (\ref{eq:GP:mean}).

The main virtue of the linear  model (\ref{eq:lineartherm}) lies in its simplicity of implementation and interpretability. In contrast, the non-parametric GP model (\ref{eq:gpthermal}) offers more modeling flexibility
and can capture
the nonlinear relation and avoid the bias of model class selection, in the cost of computational complexity. Besides, other suitable regression methods can be applied to model the thermal dynamics as well.
The choice of model depends on the practical DR requirements on computational efficiency and modeling accuracy.
The historical data above refer to the available datasets that have been collected by advanced meters before the DR event,  thus the thermal dynamics model can be estimated in an offline fashion.  Nevertheless,  dynamic regression  that uses the latest data to fit an updated model along the DR event is also applicable to further enhance the prediction accuracy.

\subsection{Learning for Customer Opt-Out Behaviors}

 Since the customer opt-out status  $z_{i,t}$ is binary,  logistic regression \cite{logis} is used to model the transition probability function $p_i(\bw_{i,t})$ in (\ref{eq:condi:1}). Because the output of logistic regression  is naturally a probability value within  $[0,1]$, and it is 
easy to implement and interpret. 
Moreover, logistic regression  is compatible with the online Bayesian learning framework with efficient posterior update approaches (see Section \ref{sec:disalg} for details). Accordingly, we formulate the transition probability function $p_i(\bw_{i,t})$ as the logistic model (\ref{eq:logist}):
\begin{align} \label{eq:logist}
 p_i(\bw_{i,t}) = \frac{1}{1+\exp( -(\alpha_i + \bm{\beta}_i^\top \bw_{i,t}) )},
\end{align}
where 
$\bm{\beta}_i$  is the weight vector describing how customer $i$ reacts to the environmental factors $\bw_{i,t}$, and $\alpha_i$ depicts the individual preference. 
Define 
$\hat{\bw}_{i,t}:=(1, \bw_{i,t})$  and $\bth_i: = (\alpha_i,\bm{\beta}_i)$.
Then the linear term in (\ref{eq:logist}) becomes $-\hat{\bw}_{i,t}^\top \bth_i$. 
Without causing any confusion, we use  $p_{\bth_i}(\bw_{i,t})$ and $p_i(\bw_{i,t})$ interchangeably.

As a consequence,  the unknown information of customer $i$'s behaviors  is summarized as vector $\bth_i$, which can be estimated from the observations of  $(\hat{\bw}_{i,t}, z_{i,t})$ in DR events. 
In contrast to the thermal dynamics model learning, the observation data of  $(\hat{\bw}_{i,t}, z_{i,t})$ are not historically available but can only be obtained along with the real implementation of DR events. This leads to an \emph{online} customer behavior learning and AC power control problem. Thus we employ the TS framework to develop the online AC control algorithm
in Section \ref{sec:algorithm} to effectively balance exploration and exploitation.

\subsection{System-Level Optimal AC Control Models}

In a typical DR setting, once a customer opts out, the AC unit will 
 automatically be switched back to the defaulted 
  operation mode with the original customer-set temperature $s_i^{\mathrm{set}}$. 
 Taking the opt-out status into account, the 
  actual AC power consumption $  \hat{u}_{i,t}$ can be formulated as
 \begin{align}\label{eq:realacp}
     \hat{u}_{i,t}= u_{i,t}\cdot  z_{i,t-1} + u_{i,t}^{\mathrm{set}}\cdot (1-z_{i,t-1}),
 \end{align}
 which equals $u_{i,t}$ if customer $i$ stays in ($z_{i,t-\!1} \!=\! 1$) or $u_{i,t}^{\mathrm{set}}$ if   opts out ($z_{i,t-\!1} \!=\! 0$). Denote
 $\bm{u}_i\!:=\!(u_{i,t})_{t\in[T]}$, $\bm{s}_i\!:=\!(s_{i,t})_{t\in[T]}$,  $\bm{d}_i\!:=\!(d_{i,t})_{t\in[T]}$, $\bm{r}_i\!:=\!(r_{i,t})_{t\in[T]}$, and $\bz:= (\bz_i)_{i\in[N]}$.
 
 To simplify the expression, we reformulate the AC control constraints \eqref{eq:ucon}, the 
 opt-out status transition \eqref{eq:condi}, \eqref{eq:wdef}, \eqref{eq:logist},  and the dynamics of environmental factors \eqref{eq:discom}, \eqref{eq:reward},   \eqref{eq:lineartherm} or \eqref{eq:gpthermal}, for customer $i\in[N]$, as the following compact form \eqref{eq:compact}:
 \begin{align} \label{eq:compact}
     (\bu_i, \bm{z}_i,\bm{s}_i, \bm{d}_i, \bm{r}_i)\in \mathcal{X}_i,
 \end{align}
 where  $\mathcal{X}_i$ denotes the corresponding feasible set.
  Then  two system-level optimal control (SOC) models, i.e. (\ref{eq:ACC1}) and (\ref{eq:ACC2}), with  different DR goals 
 are established for the SA to 
 solve  optimal AC power control schemes over $N$ customers.

\textit{1) SOC-1 model} (\ref{eq:ACC1}) aims to 
reduce as much AC load  as possible in a DR event, which can be used to flatten the load peaks or mitigate reserve deficiency emergency. 
\begin{subequations} \label{eq:ACC1}
    \begin{align}
    \begin{split}
         \min_{u_{i,t}}  &\   \mathbb{E}_{\bm{z}}\Big[ \sum_{i=1}^N\big[\Delta t \sum_{t=1}^{T} \hat{u}_{i,t}  +\rho_i(1\!-z_{i,T}) \big]
  \Big]
    \end{split} \label{eq:acc1:obj}\\
    \text{s.t}.  &\   (\bu_i, \bm{z}_i,\bm{s}_i, \bm{d}_i, \bm{r}_i)\in \mathcal{X}_i,\quad    \forall i\in [N]    \label{eq:acc1:con}
\end{align}
\end{subequations}
where objective (\ref{eq:acc1:obj}) minimizes the expected total  AC energy  consumption over the DR event, plus an opt-out penalty term in the last time step. $\rho_i$ is the penalty coefficient that can be tuned to balance load reduction and  opt-out outcomes.  $\mathbb{E}_{\bm{z}}[\cdot]$ denotes the expectation that is  taken  over  the  randomness  of the customer opt-out status $\bz$. Constraint \eqref{eq:acc1:con} collects the counterparts of \eqref{eq:compact} for all customers.

\textit{2) SOC-2 model}  (\ref{eq:ACC2}) aims to closely track a regulated  power  trajectory $(L_t)_{t\in[T]}$, which is determined by the upper-level power dispatch or the DR market.
\begin{subequations} \label{eq:ACC2}
    \begin{align}
    \begin{split}
         \min_{u_{i,t}} &\   \mathbb{E}_{\bm{z}} \Big [ \sum_{t=1}^T \Delta t\big(
 \sum_{i=1}^{N} \hat{u}_{i,t}  - L_{t}\big)^2 + \sum_{i=1}^N \rho_i(1\!-z_{i,T})  \Big]
    \end{split} \label{eq:acc2:obj}\\
     \text{s.t}.  &\  (\bu_i, \bm{z}_i,\bm{s}_i, \bm{d}_i, \bm{r}_i)\in \mathcal{X}_i,\quad    \forall i\in [N].   \label{eq:acc2:con}
\end{align}
\end{subequations}
Objective (\ref{eq:acc2:obj}) minimizes the expected total squared power  tracking deviation  from the global target $L_t$, plus the same opt-out penalty term defined in \eqref{eq:acc1:obj}. Constraint \eqref{eq:acc2:con} is the same as \eqref{eq:acc1:con}.

\begin{remark}
\normalfont The penalty term $\rho_i(1-z_{i,T})$  in the objectives \eqref{eq:acc1:obj} and \eqref{eq:acc2:obj} serves as the final state cost in a finite-horizon control planning problem, which is used to restrict the last control action $u_{i,T}$. 
Without this penalty term, the last control action $u_{i,T}$ would be too radical with no regard for the opt-out outcome at time $T$ and lead to frequent final opt-out $z_{i,T}=0$. Besides, this penalty term is a useful tool for the SA to make a trade-off between  the DR goals and the customer opt-out results through adjusting the coefficient $\rho_i$. 
\end{remark}

The two SOC models \eqref{eq:ACC1} and \eqref{eq:ACC2} are indeed discrete-time finite-horizon control planning problems, which are in the form of nonconvex stochastic optimization, and the stochasticity results from the probabilistic opt-out status transition \eqref{eq:condi}. We develop the distributed solution methods for the SOC models \eqref{eq:ACC1} and \eqref{eq:ACC2} in the next section.

%% file: Algorithm.tex
\section{Distributed Solution and Algorithm Design} \label{sec:algorithm}

For the real-time AC control in a DR event,
we pursue a distributed implementation manner such that
\begin{itemize}
    \item [1)]  the control algorithm can be directly embedded in the local home electric appliances or smart phone apps;
    \item [2)]  heavy communication burdens  between the SA and households are avoided during the DR event;
    \item [3)] the private information of customers can be protected.
\end{itemize}
In this section, we propose the distributed solution methods for the SOC models (\ref{eq:ACC1}) and (\ref{eq:ACC2}), then develop the  distributed online AC control algorithm based on the TS framework.

\subsection{Distributed Solution of SOC-1 Model (\ref{eq:ACC1})}

Since the opt-out status transition of one customer is assumed to be independent of other customers in \eqref{eq:condi}, 
 objective \eqref{eq:acc1:obj} in 
the SOC-1 model has no substantial coupling among different customers.
Hence,
the SOC-1 model  (\ref{eq:ACC1}) can be equivalently  decomposed into
 $N$  local problems,
  i.e., model (\ref{eq:LACC-1}) for each customer $i\!\in\![N]$. 
\begin{subequations} \label{eq:LACC-1}
    \begin{align}
         \min_{u_{i,t}}  &\   \mathbb{E}_{\bm{z}_i}\Big[  \Delta t\sum_{t=1}^{T} \hat{u}_{i,t}  +\rho_i(1-z_{i,T}) 
  \Big] \label{eq:lcc1:obj}\\
     \text{s.t}.  &\    (\bu_i, \bm{z}_i,\bm{s}_i, \bm{d}_i, \bm{r}_i)\in \mathcal{X}_i. \label{eq:lcc1:con}
\end{align}
\end{subequations}
 The sum of  objectives \eqref{eq:lcc1:obj} over all $N$ customers is essentially objective \eqref{eq:acc1:obj} in the SOC-1 model,  and 
constraint (\ref{eq:lcc1:con}) is the individual version of (\ref{eq:acc1:con}) for customer $i$.

The local model (\ref{eq:LACC-1}) is a stochastic optimization with the expectation over  $\bm{z}_i$ in the objective. 
Since $\bm{z}_i$ follows the transition (\ref{eq:condi}) with the probability function $p_{\bth_i}(\bw_{i,t})$  (\ref{eq:logist}), we can derive the analytic form of the expectation in (\ref{eq:lcc1:obj}), which leads to expression  (\ref{eq:ana1}):
\begin{align} 
     \sum_{t=1}^T\Big[\Delta t(u_{i,t} - u_{i,t}^{\mathrm{set}}) \prod_{\tau=1}^{t-1} p_{\bth_i}(\bw_{i,\tau}) \Big] \!-\rho_i \prod_{t=1}^{T} p_{\bth_i}(\bw_{i,t}). \label{eq:ana1}
\end{align}
Expression (\ref{eq:ana1}) 
only differs from the expectation in (\ref{eq:lcc1:obj}) by a constant term $  \Delta t\sum_{t=1}^T  u_{i,t}^{\mathrm{set}} +\rho_i$, thus they are  equivalent in optimization.
See  Appendix \ref{ap:LACC} for the detailed derivation.

When $\bth_i$ is given, we can obtain the optimal AC  control schemes $\bm{u}_i^*$ for each customer $i$ via solving the local  model (\ref{eq:LACC-1}), which is implemented in a   fully distributed  manner.
 Meanwhile, the  aggregation  of $\bm{u}_i^*$ over all $N$ customers is essentially an optimal solution to the SOC-1 model (\ref{eq:ACC1}).

\subsection{Distributed Solution of SOC-2 Model (\ref{eq:ACC2})}

The objective (\ref{eq:acc2:obj}) in the SOC-2 model  has 
coupling among different customers  due to tracking a global power trajectory $(L_t)_{t\in[T]}$. To solve this problem distributedly, we introduce  a local tracking trajectory $\bm{l}_i\!:=\!(l_{i,t})_{t\in[T]}$ for each customer $i\in[N]$ with   $ \sum_{i=1}^N l_{i,t} = L_t$ for all $t\in[T]$.
 Then
we substitute $L_t$ by $ \sum_{i=1}^N l_{i,t}$ in objective (\ref{eq:acc2:obj}), so that (\ref{eq:acc2:obj}) can be  approximated\footnote{The approximation is made by dropping the term $2 \sum_{t=1}^T\sum_{i\neq j} \mathbb{E}(\hat{u}_{i,t}-  l_{i,t})\mathbb{E}(\hat{u}_{j,t}-  l_{j,t})$ from the expansion of (\ref{eq:acc2:obj}). This term is expected to be relatively small and thus neglectable.} by the decomposable form (\ref{eq:app2:obj}), which takes the form of a sum  over $N$ customers. 
As a result, the SOC-2 model \eqref{eq:ACC2} is modified as \eqref{eq:mod:ACC2}:
\begin{subequations}\label{eq:mod:ACC2}
    \begin{align} 
     \min_{u_{i,t},\, l_{i,t}\geq 0} \, &\sum_{i=1}^N \mathbb{E}_{\bm{z}_i}\!\Big[ \sum_{t=1}^T  \Delta t\big (
\hat{u}_{i,t}- l_{i,t}\big)^2  + \rho_i(1\!-\!z_{i,T}) \Big]\label{eq:app2:obj}\\
\text{s.t.}\quad &   (\bu_i, \bm{z}_i,\bm{s}_i, \bm{d}_i, \bm{r}_i)\in \mathcal{X}_i,\quad    \forall i\in [N]  \\
& \sum_{i=1}^N l_{i,t} = L_t, \quad \forall t\in[T] \label{eq:sumlocal}
\end{align}
\end{subequations}
where $l_{i,t}$ is also a decision variable.

Consequently, the only substantial coupling among customers in the modified SOC-2 model \eqref{eq:mod:ACC2} is the equality constraint (\ref{eq:sumlocal}). Therefore, we can introduce the dual variable $\bm{\lambda}\!:=\!(\lambda_t)_{t\in[T]}$ for the equality constraint (\ref{eq:sumlocal}) and employ the \emph{dual gradient algorithm} \cite{dualgra} to solve 
the modified SOC-2 model \eqref{eq:mod:ACC2} in a distributed manner. The specific distributed solution method  is presented as Algorithm \ref{alg:primaldual}.  The implementation of 
 Algorithm \ref{alg:primaldual}  needs the  two-way  communication  of   $\bl$ and $\bm{l}_i$ between the SA and every customer  in each iteration. Due to the simple structure with only one  equality coupling constraint (\ref{eq:sumlocal}), Algorithm \ref{alg:primaldual} can converge quickly with appropriate step sizes, which is verified by our simulations.

\begin{algorithm}
 \caption{Distributed Solution Algorithm.}
 \begin{algorithmic}[1]  \label{alg:primaldual}
 
   \STATE \textbf{Initialization:} Set initial dual value $(\lambda_t^0)_{t\in[T]}$, step size $\gamma$,  convergence tolerance $\epsilon$, and iteration count $k \leftarrow 0$.
  
  \STATE \textbf{Parallel Optimization}: With the broadcast dual value $(\lambda_t^k)_{t\in[T]}$,
  each customer $i$ solves the local AC control problem (\ref{eq:local2}) and obtains the optimal solution $(\bu_i^*,\bm{l}_i^*)$. 
  \begin{subequations} \label{eq:local2}
      \begin{align}
           \min_{u_{i,t},\, l_{i,t}} & \mathbb{E}_{\bm{z}_i}\!\Big[ \Delta t\sum_{t=1}^T   (
\hat{u}_{i,t}\!-\! l_{i,t})^2 \! +\! \rho_i(1\!-\!z_{i,T}) \Big]\! +\!   \sum_{t=1}^T \lambda_t^kl_{i,t}  \label{eq:local2:obj}\\
\text{s.t.} \ & (\bu_i, \bm{z}_i,\bm{s}_i, \bm{d}_i, \bm{r}_i)\!\in \mathcal{X}_i;\  l_{i,t}\geq 0,\, \forall t\in[T].
      \end{align}
  \end{subequations}
 
  \STATE \textbf{Dual Variable Update}: Each customer $i$ uploads $\bm{l}_i^*$ to the SA, and the SA updates the dual variable by
  \begin{align}
      \lambda_t^{k+1} \leftarrow \lambda_t^{k} +  \gamma\cdot (\sum_{i=1}^N l_{i,t}^* - L_t), \quad \forall t\in[T].
  \end{align}
  
  \STATE \textbf{Convergence Check}: If $||\bl^{k+1} \!-\!\bl^k||\leq \epsilon$, terminate the algorithm. Otherwise, let $k\leftarrow k+1$ and go  to Step 2. 
  
 \end{algorithmic} 
 \end{algorithm}

Similar to \eqref{eq:ana1},  we can derive the equivalent analytic form  (\ref{eq:ana2}) for 
the expectation term in (\ref{eq:local2:obj}): 
\begin{align} \label{eq:ana2}
\begin{split}
     & \Delta t\sum_{t=1}^T \bigg\{ ( u_{i,t}^{\mathrm{set}} -l_{i,t} )^2 +\Big[ 2(u_{i,t}^{\mathrm{set}}  - l_{i,t})(
u_{i,t} -u_{i,t}^{\mathrm{set}})\\
&+
(u_{i,t} -u_{i,t}^{\mathrm{set}})^2\Big]\cdot \prod_{\tau=1}^{t-1} p_{\bth_i}(\bw_{i,\tau})\bigg \} -\rho_i \prod_{t=1}^{T} p_{\bth_i}(\bw_{i,t}).
\end{split}
\end{align} 
By substituting the expectations terms in \eqref{eq:lcc1:obj} and \eqref{eq:local2:obj} with their 
 analytic forms (\ref{eq:ana1}) and (\ref{eq:ana2}) respectively, the local AC control models (\ref{eq:LACC-1}) and (\ref{eq:local2}) 
 become  deterministic nonconvex optimization problems. Given parameter $\bth_i$, they can be solved efficiently via available nonlinear optimizer tools, such as  the IPOPT solver \cite{ipopt}.
 For concise expression, we denote 
the above distributed solution methods together with the  optimizer  tools as an  oracle 
\begin{align}\label{eq:oracle}
    \mathcal{O}:\bm{\theta}_i \rightarrow \bm{u}_i^*,
\end{align}
which generates optimal $\bm{u}_i^*$ with the input of parameter $\bth_i$.

\subsection{Distributed  Online DR Control Algorithm} \label{sec:disalg}

\begin{figure}[thpb]
	\centering
		\includegraphics[scale=0.47]{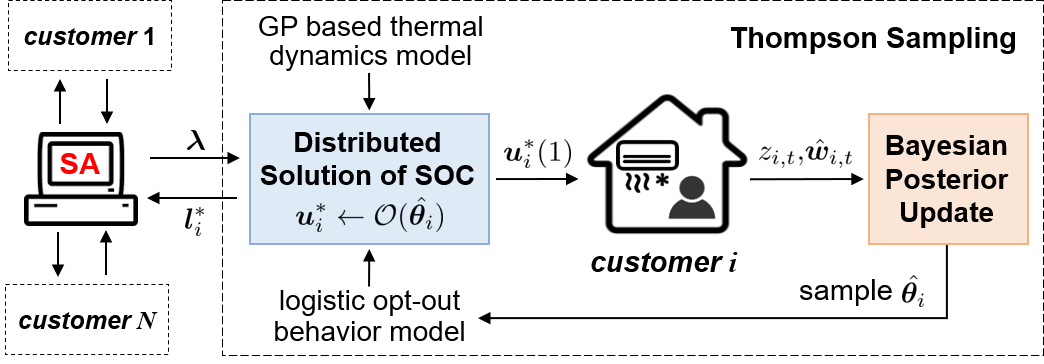}
    	\caption{Schematic of the distributed online DR control algorithm implemented on the SOC-2 problem. (For the SOC-1 problem, the communications of $\bm{\lambda}$ and $\bm{l}^*_i$ are not required. )	}
	\label{fig:frame}
\end{figure}

Based on the TS framework, we develop 
the distributed online DR control algorithm as Algorithm \ref{onlinealg} to learn customer opt-out behaviors and optimally control the AC power in a DR event.  Since this online algorithm is implemented distributedly, we present 
Algorithm \ref{onlinealg} from the perspective of an individual customer $i\in[N]$. The practical implementation of Algorithm \ref{onlinealg} is 
 illustrated as Figure \ref{fig:frame}. 
 
 \begin{algorithm} 
 \caption{Distributed Online DR Control Algorithm.}
 \begin{algorithmic}[1]  \label{onlinealg}
 \renewcommand{\algorithmicrequire}{\textbf{Input:}}
 \renewcommand{\algorithmicensure}{\textbf{Output:}}

  \STATE \textbf{Input:} Prior distribution $\sN(\bmu_i, \bm{\Sigma}_i)$. 
  Receive the DR objective and associated parameters from the SA. 
  
  \FOR {$t = 1$ to $T$}
 
  \STATE \textbf{Sample} $\hat{\bth}_i\sim \mathcal{N}\left( {\bm{\mu}_i}, {\bm{\Sigma}_i}   \right)$.

 \STATE  \textbf{Optimization}. Solve the SOC model (\ref{eq:ACC1}) or (\ref{eq:mod:ACC2}) distributedly
  for the  remaining time horizon $\{t, \cdots,T\}$, and obtain the optimal AC power control scheme using the  oracle $\bu^*_i\leftarrow \mathcal{O}(\hat{\bth}_i)$ (\ref{eq:oracle}).
 
  \STATE \textbf{Action}. Implement the first control action $\bu_i^*(1)$, and observe the customer opt-out outcome $z_{i,t}$. Collect the environmental factors $\hat{\bw}_{i,t}$ at time $t$.
  
  \STATE \textbf{Posterior Update}. \label{st:post}
 Initialize 
variational parameter $\xi_i$ by
\begin{align}\label{eq:post1}
    \xi_i \leftarrow  \sqrt{{\hat{\bw}}_{i,t}^\top{\bm{\Sigma}_i}{\hat{\bw}}_{i,t} +({\hat{\bw}}_{i,t}^\top {\bmu}_i)^2  }.
\end{align}
        
  Iterate three times between the posterior update
     \begin{align} \label{eq:post2}
    & \begin{cases}
       \bm{\hat{\Sigma}}_i^{-1}  \leftarrow {\bm{\Sigma}}_i^{-1}+2|\ell(\xi_i)|  {\hat{\bw}}_{i,t}{\hat{\bw}}_{i,t}^\top,\\
       \   \bm{\hat{\mu}}_i  \    \, \leftarrow  \bm{\hat{\Sigma}}_i\left[ {\bm{\Sigma}_i^{-1}}\bmu_i +(z_{i,t}-\frac{1}{2}){\hat{\bw}}_{i,t}\right],
     \end{cases}\\
     &\  \text{where}\ \ell(\xi_i): = (1/2-1/(1+e^{-\xi_i}))/2\xi_i. \nonumber
   \end{align}
  and  the $\xi_i$ update 
  \begin{align} \label{eq:post3}
       \xi_i\ \leftarrow  \sqrt{{\hat{\bw}}_{i,t}^\top{\bm{\hat{\Sigma}}}_i{\hat{\bw}}_{i,t} +({\hat{\bw}}_{i,t}^\top \bm{\hat{\mu}}_i)^2  }.\ \
  \end{align}
Then set
 ${\bm{\Sigma}}_i \  \leftarrow   \bm{\hat{\Sigma}}_i,\ \mathcal{{\bm{\mu}}}_i\ \leftarrow \bm{\hat{\mu}}_i.$

\STATE \textbf{Check Termination}. If $z_{i,t} =0$, terminate the AC control and change the AC operation mode back to the original customized setting.
 
  \ENDFOR
 \end{algorithmic} 
 \end{algorithm}

  In Algorithm \ref{onlinealg}, the unknown customer behavior parameter $\bth_i$  is treated as a random variable, and we  construct a Gaussian prior distribution  $\sN(\bmu_i, \bm{\Sigma}_i)$ for it based on  historical information. At each time $t\in[T]$ of a DR event,  $\hat{\bth}_i$ is randomly sampled from the distribution for decision-making. Two key techniques used in  Algorithm \ref{onlinealg} are explained as follows.

1)  To utilize the latest information  and 
take future time-slots into account, we employ the model predictive control (MPC)  method in the optimization and action steps. 
Specifically, at each time $t$,  it solves the SOC model (\ref{eq:ACC1}) or (\ref{eq:ACC2}) for the rest of the DR event to obtain the optimal AC control scheme $\bm{u}_i^*$,
 but only implements the first control action $\bm{u}_i^*$(1).  In addition, the latest predictions or estimations of the environmental factors, the updated thermal dynamics model, and recalculated baseline AC power $u_{i,t}^\mathrm{set}$ can be adopted in the optimization step.


2) After observing the outcome pair $(z_{i,t},\hat{\bw}_{i,t})$, 
the \emph{variational Bayesian inference approach} introduced in \cite{varia} is applied to obtain the posterior distribution on $\bth_i$ with the update rules (\ref{eq:post1})-(\ref{eq:post3}). It is well known that Bayesian inference for the logistic regression model (\ref{eq:logist}) is  an intrinsically hard problem \cite{pg1}, and the exact posterior $\mathcal{P}(\bth_i| z_{i,t},\hat{\bw}_{i,t})$ is intractable to compute. Thus we use the variational
 approach  \cite{varia} for efficient  Bayesian inference, which  provides an accurate Gaussian approximation to the exact posterior with a closed form \eqref{eq:post2}.  The scalar $\xi_i$ in (\ref{eq:post1})-(\ref{eq:post3}) is an intermediate  parameter that affects the approximation accuracy, thus we alternate three times between the posterior update (\ref{eq:post2}) and the $\xi_i$ update (\ref{eq:post3}), which leads to an optimal approximated posterior.
 See reference \cite{varia} for details.

\subsection{Performance Measurement} \label{sec:regret}

For online learning  problems, the notion of ``regret" and its variants are standard metrics that are
defined to 
 measure the  performance of  online learning and decision algorithms \cite{regdef}. Accordingly, we 
 denote $\bth^\star:=(\bth^\star_i)_{i\in[N]}$ as the underlying true customer behavior parameter that the LSEs do not know but aim to learn. Then
  the regret of the proposed online DR control algorithm
at $m$-th DR event is defined as
\begin{align} \label{eq:regret}
    \mathrm{regret}(m) :=  C^{\bth^\star}_m(\bm{u}^{\mathrm{online}}_m)-  C^{\bth^\star}_m(\bm{u}^{\star}_m),
\end{align}
where $C^{\bth^*}_m(\cdot)$ denotes the objective function 
(i.e., \eqref{eq:acc1:obj} in the SOC-1 model or \eqref{eq:app2:obj} in the modified SOC-2 model) under the true value $\bth^\star$. $\bm{u}^{\mathrm{online}}_m$ denotes the AC control scheme generated by the proposed online  algorithm, while $\bm{u}^{\star}_m$ is the optimal AC control scheme that minimizes the objective $C^{\bth^*}_m(\cdot)$. Thus $\mathrm{regret}(m)$ in \eqref{eq:regret} is always non-negative and measures  the performance distance between the proposed online  algorithm and the underlying best  control scheme. 

To evaluate the overall learning performance, we further define the  cumulative regret until $M$-th DR event as
\begin{align}
    \mathrm{cumu\_regret}(M):= \sum_{m=1}^M \text{regret}(m),
\end{align}
which is simply the sum of $\mathrm{regret}(m)$ over the first $M$ DR events.
Generally, a sublinear cumulative regret over  $M$ is desired, i.e., $\mathrm{cumu\_regret}(M)/M\to 0$ as $M\to +\infty$, since it indicates that  $\mathrm{regret}(m)\to 0$ as $m\to +\infty$. In other word, it means that
the proposed online algorithm can eventually learn the customer behaviors well and make optimal AC control schemes
as more and more DR events are experienced. We demonstrate the regret results of the proposed algorithm  via numerical simulations in the next section.

%% file: Simulation.tex
\section{Numerical Simulations} \label{sec:simulation}

In this section, we first test the performance of the linear thermal dynamics model and the GP model. Then, we implement the proposed distributed online DR control algorithm on the two SOC models.

\subsection{Indoor Thermal Dynamics Prediction}

In this part, we compare the thermal dynamics prediction performance of the linear model (\ref{eq:lineartherm}) and the GP model (\ref{eq:gpthermal}). Real customer metering data, including indoor temperature and AC power, from ThinkEco\footnote{ThinkEco Inc. is a New York City-based energy efficiency and demand response solution company (\url{http://www.thinkecoinc.com/}). }
 are used for model training and testing. The outdoor temperature data are procured from an open-access meteorological database\footnote{Iowa Environmental Mesonet [online]: \url{https://mesonet.agron.iastate.edu/request/download.phtml?network=MA_ASOS}.}.  Specifically, we use
the time series of data in consecutive 5 days 
with the resolution of 15 minutes to fit the two thermal dynamics models (\ref{eq:lineartherm})  and (\ref{eq:gpthermal}). The GPML  toolbox  \cite{gpml} is applied to implement the GP model  and optimize the hyperparameters. Then, the fitted models are tested on the  time series of real data in the next 3 days for indoor temperature prediction. \begin{figure}[thpb]
	\centering
	\includegraphics[scale=0.3]{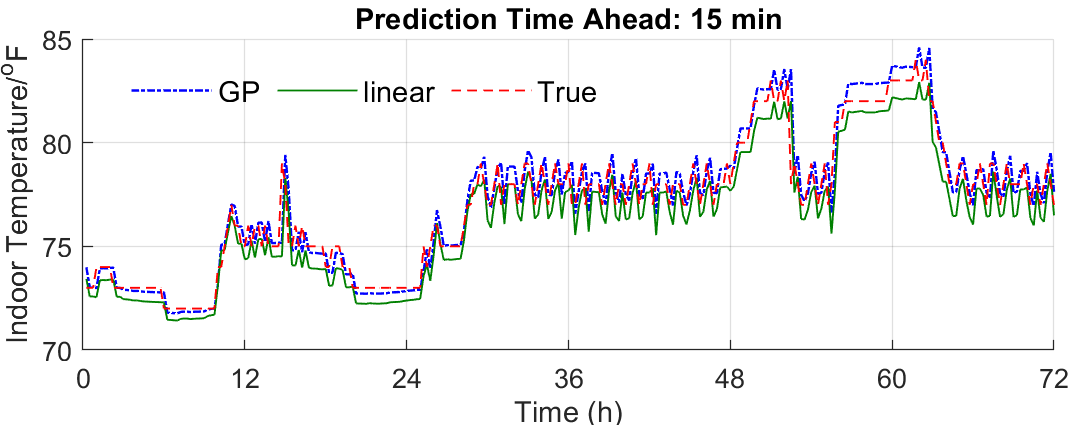}
		\includegraphics[scale=0.3]{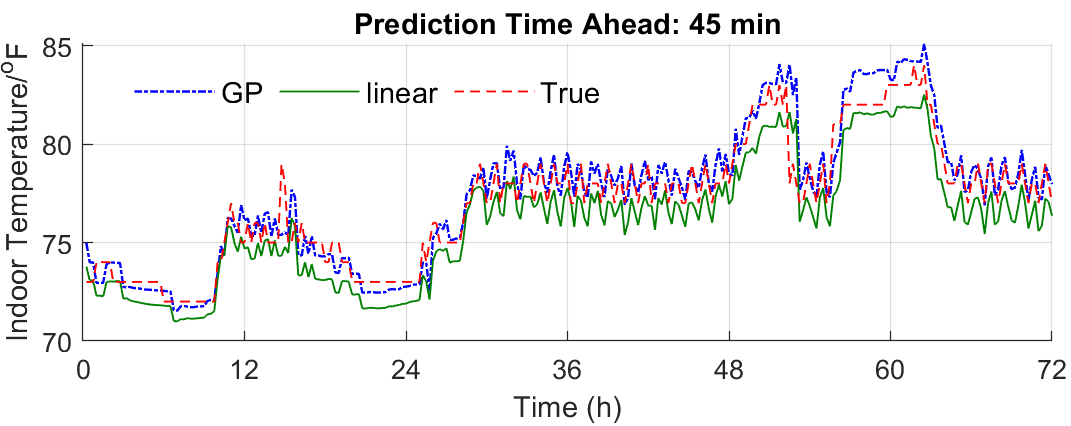}
	\caption{Indoor temperature prediction results with  Gaussian process and  linear thermal dynamics models. (GP model: blue dotted curve; Linear model: green curve; True indoor temperature: red dashed curve.)}
	\label{fig:thermal}
\end{figure}

The prediction results of one time step ahead (15 minutes) and three time steps ahead (45 minutes) are presented in Figure \ref{fig:thermal}. The average prediction errors of the indoor temperature are $0.632^{\circ}F$ and $0.737^{\circ}F$ for the GP model  in these two cases, and $0.846^{\circ}F$ and $1.16^{\circ}F$ for the linear model. It is observed  that 
both the GP model \eqref{eq:gpthermal} and the linear model \eqref{eq:lineartherm} work well in the thermal dynamics modelling, and the GP model 
achieves better prediction accuracy.

\subsection{Learning and AC Control with SOC-1 Model } \label{sec:sim:1}

\subsubsection{Simulation Configuration}

Each DR event lasts for 3 hours with the AC control period $\Delta t = 15$ minutes, which implies a time length $T = 12$. 
The AC capacity and drift limit are set  as $\bar{u}_i = 2$kW and $\Delta u_i =1$kW, and the AC setting temperature $s_i^{\text{set}}$ is $72^\circ F$.
As defined in Section \ref{sec:regret}, 
we associate each customer $i\in[N]$ with a true behavior parameter $\bth_i^{\star}\in \mathbb{R}^6$ to simulate the opt-out outcomes,  whose values are randomly generated but satisfy several basic rules to be reasonable.
For example, if no DR control is implemented (defaulted AC setting), the stay-in probability $g_{\bth_i^{\star}}(\bw_{i,t})$ should be very close to 1; if the indoor temperature $s_{i,t}$ reaches a high value such as $90^\circ F$,  the stay-in probability  should be very close to 0.  
The considered environmental factors include indoor temperature $s_{i,t}$, accumulated thermal discomfort $d_{i,t}$, incentive credit $r_{i,t}$, outdoor temperature $s_{i,t}^{\mathrm{out}}$, and time-varying electricity price, where the first three factors follow the dynamics \eqref{eq:therm}-\eqref{eq:reward} respectively, while the electricity price at each time is normalized and randomly generated from $\mathrm{Unif}(0,1)$.
Besides,
 IPOPT solver  \cite{ipopt} is employed to solve the nonconvex optimal AC control models.

\subsubsection{Control and Learning Performance}

\begin{figure}[thpb]
	\centering
	\includegraphics[scale=0.28]{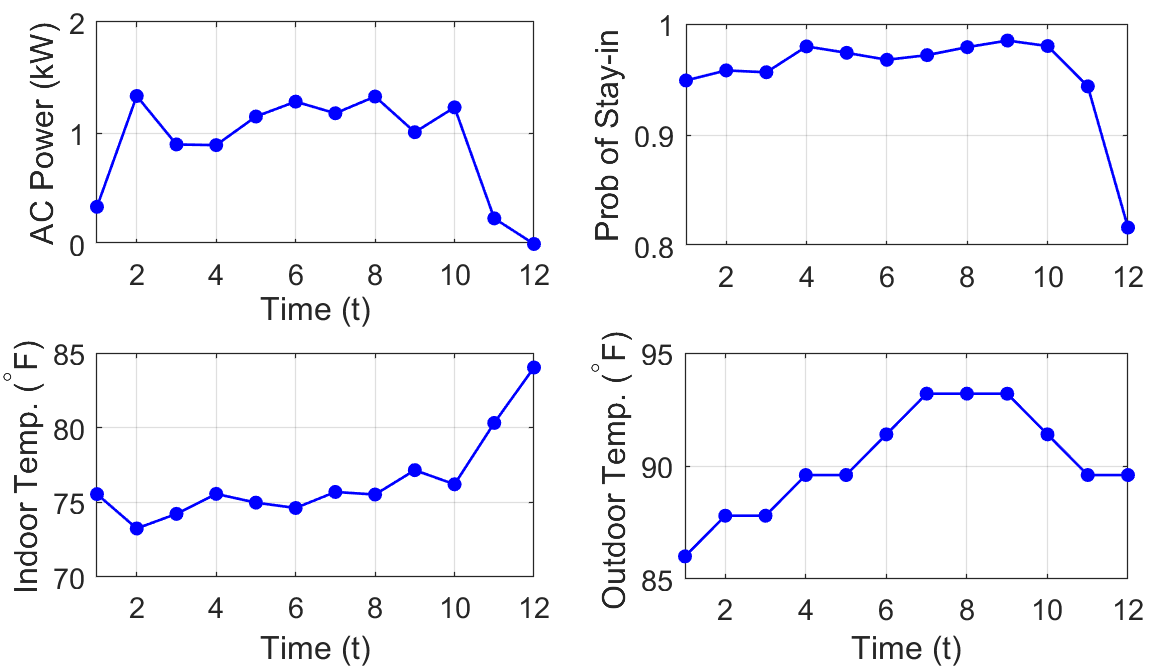}
	\caption{Illustration of the optimal AC power control scheme $\bu_i^\star$, stay-in probability $p_{\bth_i^\star}(\bw_{i,t})$, indoor temperature $s_{i,t}$, outdoor temperature $s_{i,t}^{\mathrm{out}}$
	for a customer, which are computed via solving model (\ref{eq:LACC-1}) with known $\bth_i^{\star} $.}
	\label{fig:ACscheme1}
\end{figure}

Since the aggregation of all local optimal AC control schemes is an optimal solution to the SOC-1 model (\ref{eq:ACC1}), we simulate the AC control and learning for a single customer over sequential DR events. Given the true parameter $\bth_i^{\star}$, the optimal AC power trajectory $\bu_i^{\star}$ can be computed via
 solving the local control model (\ref{eq:LACC-1}).  Figure \ref{fig:ACscheme1} illustrates the simulation results associated with the  optimal AC control scheme $\bu_i^{\star}$.
 It is seen that
the stay-in probability is maintained close to 1 by the AC control scheme, which tends to make customers comfortable and not opt out for the sake of  long-term load reduction. Besides, there is a  drop of AC power at the end of the DR event ($t=11,12$), leading to increased indoor temperature and a  decrease in the stay-in probability. Intuitively, that is because last-minute opt-out will not affect the DR objective much, and thus  a radical AC power reduction is conducted. This effect can be mitigated by increasing the penalty coefficient $\rho_i$.

\begin{figure}[thpb]
	\centering
	\includegraphics[scale=0.28]{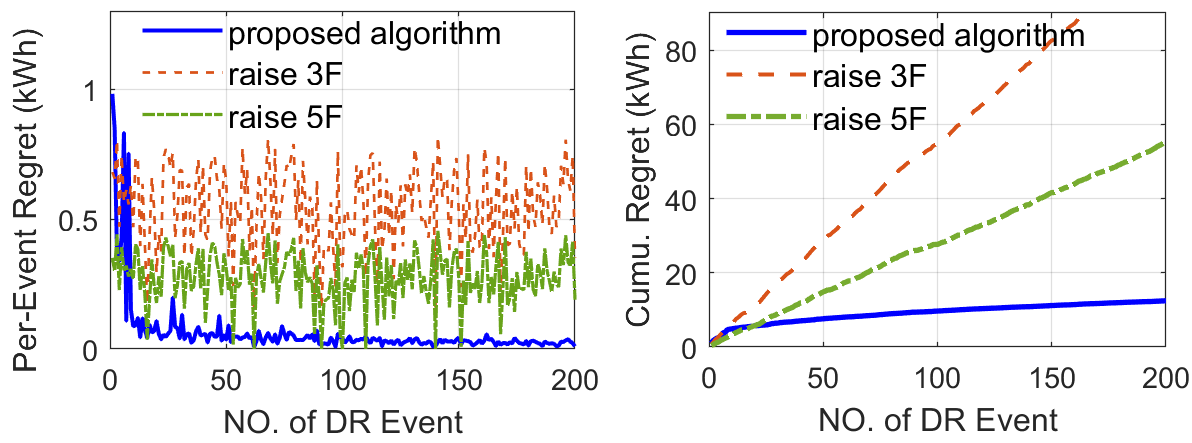}
	\caption{The regret comparison between the proposed  Algorithm \ref{onlinealg} and two baseline schemes (raising the AC setting temperature by $3^\circ F$ and $5^\circ F$) for a typical  customer  under the SOC-1 model (\ref{eq:ACC1}).}
	\label{fig:regret1}
\end{figure}

In practice, the true customer parameter $\bth_i^{\star}$ is unknown, and we implement the proposed online algorithm  to learn  customer behaviors and make AC control decisions in sequential DR events. We compare the performance of the proposed algorithm with two baseline schemes:  raising the AC setting temperature by $3^\circ F$ and   $5^\circ F$ respectively, which are typically used in real DR programs.
The regret results for one customer over 200 DR events are shown as Figure \ref{fig:regret1}. It is observed that the per-event $\text{regret}(m)$ of the proposed algorithm decreases dramatically within the first tens of DR events, then converges to almost zero. As a result, 
the associated cumulative regret exhibits a clear sublinear  trend, which verifies the learning efficiency of the proposed online algorithm. In contrast, the baseline schemes that simply raise the AC setting temperature without consideration of customer opt-out behaviors maintain high regret values. Besides, the average amount of load energy reduction  of one customer in a DR event is $1.13 \text{ kWh}$
for the proposed algorithm, which is higher than the baseline schemes with $0.6 \text{ kWh}$ and $0.82 \text{ kWh}$  load reduction, respectively.

\subsection{Learning and AC Control with SOC-2 Model }

We then perform the proposed online DR control algorithm on  the  SOC-2 model, where the simulation configurations are the same as Section \ref{sec:sim:1}. Since the SOC-2 problem involves the coordination among all customers, we set 
the customer number as $N=500$, and the global tracking target  $L_t$ is randomly generated from $\text{Unif}(450, 550)$(kW).

\subsubsection{Distributed Solution}

\begin{figure}[thpb]
	\centering
		\includegraphics[scale=0.32]{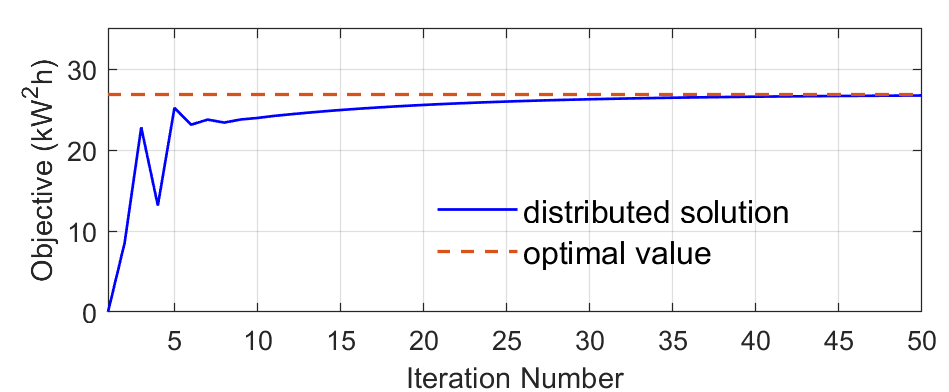}
	\caption{Convergence results of the distributed algorithm (Algorithm \ref{alg:primaldual}) for solving the modified SOC-2 model \eqref{eq:mod:ACC2}.}
	\label{fig:converge}
\end{figure}

\begin{figure}[thpb]
	\centering
		\includegraphics[scale=0.28]{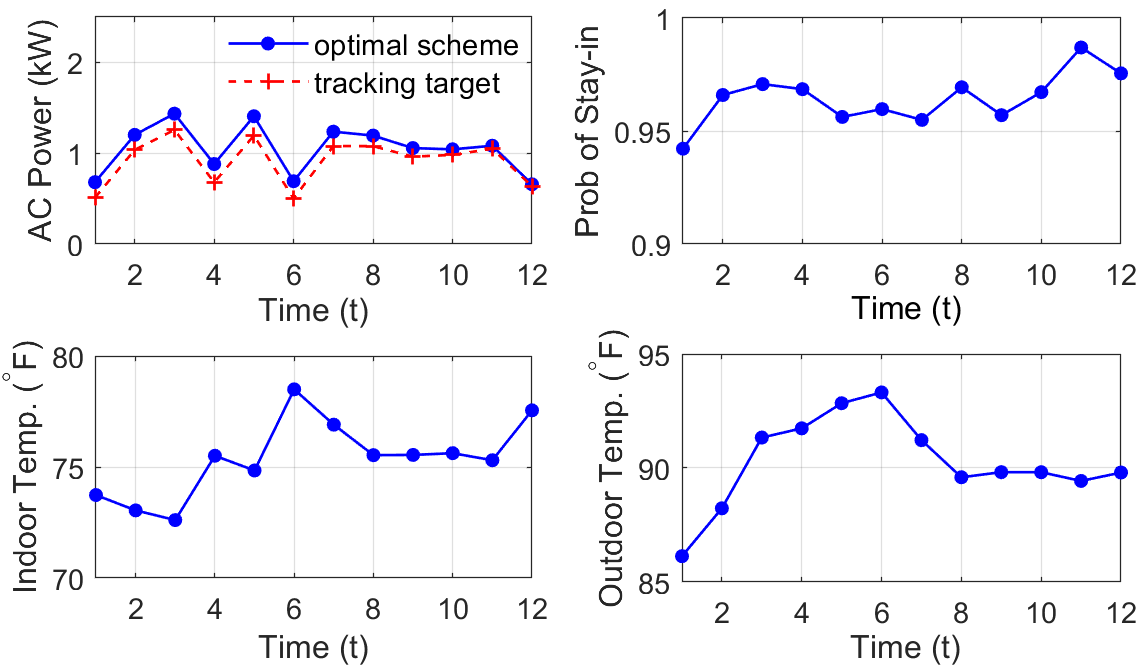}
\caption{Illustration of the optimal AC power control scheme $\bu_i^\star$ with the local tracking target $\bm{l}_i^\star$, stay-in probability $p_{\bth_i^\star}(\bw_{i,t})$, indoor temperature $s_{i,t}$, outdoor temperature $s_{i,t}^{\mathrm{out}}$
	for a customer, which are computed via  solving the modified SOC-2 model \eqref{eq:mod:ACC2} with known $\bth_i^{\star} $.}		
	\label{fig:known2}
\end{figure}

We apply Algorithm \ref{alg:primaldual} to
solve the modified SOC-2 model \eqref{eq:mod:ACC2} distributedly over 500 customers. A diminishing step size with $\gamma_k \!= \!\max(5\times 10^{-4}/\sqrt{k}, 1\times 10^{-4})$ is employed to speed up the convergence. For the case with given  $\bth_i^{\star}$, the convergence results are shown as Figure \ref{fig:converge}. It is seen that the distributed algorithm converges to the optimal value  within tens of iterations.
We note that the convergence curve  in Figure \ref{fig:converge} and the associated AC power are just intermediate computational  values, which are not executed in practice, while only the final converged values are regarded as the AC power control scheme and are
 implemented.  Figure \ref{fig:known2} illustrates the simulation results associated with the converged optimal AC control scheme, including the optimal AC power with the local tracking target, the stay-in probability, and the indoor/outdoor temperature profiles  for one  customer. 

\subsubsection{Learning Performance}

\begin{figure}[thpb]
	\centering
	\includegraphics[scale=0.22]{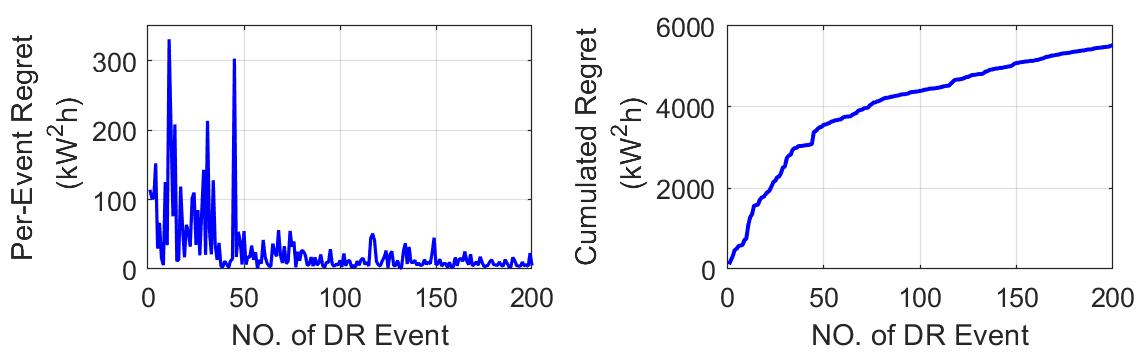}
	\caption{The regret results of the proposed Algorithm \ref{onlinealg} on the modified SOC-2 model \eqref{eq:mod:ACC2} with  500 customers.}
	\label{fig:regret2}
\end{figure}

In practice, the true parameter $\bth_i^{\star}$ is unknown,   therefore we 
implement the proposed online algorithm to learn customer opt-out behaviors and make AC control decisions with  the modified SOC-2 model \eqref{eq:mod:ACC2}. 
 The regret results of the proposed algorithm for 200 DR events over 500 customers are  presented in Figure \ref{fig:regret2}. 
A rapidly decreasing  regret and a sublinear cumulative regret are observed, which verify that the proposed algorithm can learn the customer behaviors well and generate efficient AC control decisions.  Besides,  for the per-event regret curve in Figure \ref{fig:regret2},   
 non-monotonic variations and 
  occasional spikes are observed in the early  learning stage (similar in Figure \ref{fig:regret1}). That is because the proposed algorithm follows the TS framework and draws a random sample $\hat{\bth_i}$ for decision-making at each time, which 
 results in the non-monotonic variations with stochasticity. In addition, 
there is a small chance that the sample $\hat{\bth}_i$ is quite different from the true value $\bth^\star_i$, leading to large spikes  in the regret curve. However, 
as the observed customer opt-out outcomes accumulate, the distribution on $\bth_i$ gradually concentrates on the true value $\bth_i^\star$, so that there are less large spikes in the later learning stage.

%% file: Conclusion.tex
\section{Conclusion}\label{sec:conclusion}

In this paper, we propose a distributed online DR control algorithm to learn customer  behaviors and regulate AC loads for incentive-based residential DR. Based on the Thompson sampling framework, the proposed algorithm consists of an online Bayesian learning step and an offline distributed optimization step. Two DR objectives, i.e. minimizing total AC loads and closely tracking a regulated power trajectory, are considered. 
The numerical simulations show that the distributed solution  converges to the optimal value within tens of iterations, and 
the  regret of learning reduces rapidly on average along with the implementation of DR events. Future works include 1) identify significant and effective environmental factors based on real user data; 2)  conduct practical DR experiments using the proposed algorithm and  analyze its practical performance.

%% file: Appendix1.tex
\appendices

\section{ Analytic Derivation for Expectation in (\ref{eq:lcc1:obj}) }\label{ap:LACC}

The expectation term in (\ref{eq:lcc1:obj}) can be expanded as 
\begin{align*}
  \mathbb{E}_{\bm{z}_i}\bigg[
  \sum_{t=1}^{T}\!\Delta t \left( (u_{i,t}- u_{i,t}^{\mathrm{set}}) z_{i,t-1}\right)\!-\rho_iz_{i,T} 
  \bigg] \!+ \Delta t\sum_{t=1}^T u_{i,t}^{\mathrm{set}} +\rho_i.
\end{align*}
The  status transition (\ref{eq:condi}) implies that
 the  customer opt-out time ${t}_i^{\mathrm{opt}}$ actually follows a geometric distribution with the Bernoulli probability $p_{i,t}\!:=\!p_{\bth_i}(\bw_{i,t})$ (\ref{eq:logist}). 
 Thus,  we can enumerate all the 
 possible realizations with different
 ${t}_i^{\mathrm{opt}}$ and the probabilities:
 \begin{align*}
     \begin{cases}
    \Delta t(u_{i,1} - u_{i,1}^{\mathrm{set}}), & \mathbb{P}({t}_i^{\mathrm{opt}}\!=\!1)= 1 - p_{i,1}\\
 \cdots  & \cdots \\
   \Delta t\!\sum_{t=1}^{T}\!  (u_{i,t}\! -\! u_{i,t}^{\mathrm{set}}), & \mathbb{P}({t}_i^{\mathrm{opt}}\!=\!T)\!=\! (1\!-\!p_{i,T})\prod_{t=1}^{T\!-\!1} p_{i,t} \\
   \Delta t\! \sum_{t=1}^{T}\!  (u_{i,t}\! -\! u_{i,t}^{\mathrm{set}})\!-\!\rho_i, & \mathbb{P}({t}_i^{\mathrm{opt}}\!=\! \text{not})\!=\! \prod_{t=1}^{T} p_{i,t}
     \end{cases}
 \end{align*}
where ``${t}_i^{\mathrm{opt}}\!=\! \text{not}$" means  customer $i$ does not opt out at any time. Then, the expectation term is computed analytically by summing up all these cases, which leads to (\ref{eq:ana1}) plus the constant term  $\Delta t\sum_{t=1}^T u_{i,t}^{\mathrm{set}} +\rho_i$.